\documentclass[11pt]{amsart}
\usepackage{amsmath,amsfonts,amssymb,amsthm,graphicx,euscript,mathrsfs}
\IfFileExists{srcltx.sty}{\RequirePackage[active]{srcltx}}{}

\newcommand{\Sob}{\EuScript{H}}

\usepackage[numbers]{natbib}
\usepackage{xy}
    \xyoption{matrix}
    \xyoption{arrow}

\newtheorem{assumption}{Assumption}
\newtheorem{theorem}{Theorem}
\newtheorem{lemma}[theorem]{Lemma}
\newtheorem{corollary}[theorem]{Corollary}

\theoremstyle{definition}
\newtheorem{definition}[theorem]{Definition}

\theoremstyle{remark}

    \numberwithin{equation}{section}
    \numberwithin{theorem}{section}

\newcommand{\cD}{\mathcal{ D}}
\newcommand{\cE}{\mathcal{ E}}
\newcommand{\cF}{\mathcal{ F}}
\newcommand{\cG}{\mathcal{ G}}
\newcommand{\cH}{\mathcal{ H}}

\newcommand{\cL}{\mathcal{ L}}

\newcommand{\cN}{\mathcal{ N}}
\newcommand{\cO}{\mathcal{ O}}

\newcommand{\cS}{\mathcal{ S}}
\newcommand{\cT}{\mathcal{ T}}

\newcommand{\C}{\mathbb{C}}
\newcommand{\Z}{\mathbb{Z}}

\def\R{\mathbb{R}}

\def\bH{{\mathbb{H}^{\,}}}

\def\g{\mathfrak{g}}

\def\th{\theta}
\def\vth{\vartheta}

\def\la{\lambda}
\def\Om{\Omega}

\def\al{\alpha}
\def\Th{\Theta}
\def\Ga{\Gamma}
\def\de{\delta}
\def\De{\Delta}

\title{Quantum Field Theory on Curved Backgrounds. II.\\
Spacetime Symmetries}

\author{Arthur Jaffe}
\email{arthur\_jaffe@harvard.edu}
\address{Harvard University\\
17 Oxford St.\\
Cambridge, MA 02138}

\author{Gordon Ritter}
\email{ritter@post.harvard.edu}

\address{Harvard University\\
17 Oxford St.\\
Cambridge, MA 02138}

\date{February 22, 2007}

\def\mbni#1{\medbreak \noindent {\bf #1}\enspace}

\newcommand{\lrp}[1]{\left( #1 \right)}

\newcommand{\lra}[1]{\left\langle #1 \right\rangle}
\newcommand{\lrabig}[1]{\big\langle #1 \big\rangle}

\newcommand{\abs}[1]{\left| #1 \right|}
\newcommand{\norm}[1]{{\| #1 \|}}

\renewcommand{\d}{\partial}

\def\<{\langle}
\def\>{\rangle}

\def\hook{{\font\xx=msam10 \hbox{\xx \char22}}}

\newcommand\supp{\operatorname{supp}}

\newcommand\Iso{\operatorname{Iso}}
\newcommand\Isom{\operatorname{Isom}}

\newcommand{\Lie}{\operatorname{Lie}}

\newcommand{\SO}{\operatorname{SO}}

\newcommand\te{\text}

\newcommand\f{\frac}

\newcommand\bea{\begin{eqnarray}}
\newcommand\eea{\end{eqnarray}}
\newcommand\beas{\begin{eqnarray*}}
\newcommand\eeas{\end{eqnarray*}}

\newcommand{\er}[1]{\eqref{#1}}
\newcommand{\bel}[1]{\begin{equation}\label{#1}}
\newcommand\ee{\end{equation}}

\newcommand\nn{\nonumber}

\newcommand{\cale}{\mathcal{E}}

\newfont{\boldit}{cmbxti10}

\newcommand{\Vect}{\operatorname{Vect}}

\newcommand{\Lor}[1]{{#1}_{\text{lor}}}
\newcommand{\Dhat}{\widehat{\cD}}

\def\mathcal{\mathscr}

\begin{document}

\begin{abstract}
We study space-time symmetries in scalar quantum field theory
(including interacting theories) on static space-times. We first
consider Euclidean quantum field theory on a static Riemannian
manifold, and show that the isometry group is generated by
one-parameter subgroups which have either self-adjoint or unitary
quantizations. We analytically continue the self-adjoint
semigroups to one-parameter unitary groups, and thus construct a
unitary representation of the isometry group of the associated
Lorentzian manifold. The method is illustrated for the example of
hyperbolic space, whose Lorentzian continuation is Anti-de Sitter
space.
\end{abstract}

\maketitle

\section{Introduction}

The extension of quantum field theory to curved space-times has led to the
discovery of many qualitatively new phenomena which do not occur in the simpler
theory on Minkowski space, such as Hawking radiation; for background and
historical references, see \cite{BirrellDavies,Fulling,WaldSurvey}.

The reconstruction of quantum field theory on a Lorentz-signature space-time
from the corresponding Euclidean quantum field theory makes use of
Osterwalder-Schrader (OS) positivity \cite{OS1,OS2} and analytic continuation.
On a curved background, there may be no proper definition of time-translation
and no Hamiltonian; thus, the mathematical framework of Euclidean quantum field
theory may break down. However, on static space-times there is a Hamiltonian
and it makes sense to define Euclidean QFT. This approach was recently taken by
the authors \cite{JR:CurvedSp}, in which the fundamental properties of
Osterwalder-Schrader quantization and some of the fundamental estimates of
constructive quantum field theory\footnote{For background on constructive field
theory in flat space-times, see \cite{GJ,Jaffe:2000a}.} were generalized to
static space-times.

The previous work \cite{JR:CurvedSp}, however, did not address the
analytic continuation which leads from a Euclidean theory to a
real-time theory. In the present article, we initiate a treatment
of the analytic continuation by constructing unitary operators
which form a representation of the isometry group of the
Lorentz-signature space-time associated to a static Riemannian
space-time. Our approach is similar in spirit to that of
Fr\"ohlich \cite{Fr:80} and of Klein and Landau \cite{KL:87}, who
showed how to go from the {E}uclidean group to the {P}oincar\'e
group without using the field operators on flat space-time.

This work also has applications to representation theory, as it
provides a natural (functorial) quantization procedure which
constructs nontrivial unitary representations of those Lie groups
which arise as isometry groups of static, Lorentz-signature
space-times. These groups are typically noncompact. For example,
when applied to $AdS_{d+1}$, our procedure gives a unitary
representation of the identity component of $SO(d,2)$. Moreover,
our procedure makes use of the Cartan decomposition, a standard
tool in representation theory.

\section{Classical Space-Time}
\subsection{Structure of Static Space-Times}\label{sec:fundamentals}

\begin{definition}\label{def:localization}
A {\bf quantizable static space-time} is a complete, connected orientable
Riemannian manifold $(M, g_{ab})$ with a globally-defined (smooth) Killing
field $\xi$ which is orthogonal to a codimension-one hypersurface $\Sigma
\subset M$, such that the orbits of $\xi$ are complete and each orbit
intersects $\Sigma$ exactly once.
\end{definition}

Throughout this paper, we assume that $M$ is a quantizable static space-time.
Definition \ref{def:localization} implies that there is a global time function
$t$ defined up to a constant by the requirement that $\xi = \d / \d t$. Thus
$M$ is foliated by time-slices $M_t$, and
\[
    M = \Om_- \cup \Sigma \cup \Om_+
\]
where the unions are disjoint, $\Sigma = M_0$, and $\Om_{\pm}$ are
open sets corresponding to $t>0$ and $t<0$ respectively. We infer existence of an isometry $\th$ which reverses the sign of $t$,
\[
    \th : \Om_\pm \to \Om_\mp\ \te{ such that } \ \th^2 = 1,
    \ \ \th|_\Sigma = \text{id} .
\]

Fix a self-adjoint extension of the Laplacian, and let $C = (- \De
+ m^2)^{-1}$ be the resolvent of the Laplacian (also called the
\emph{free covariance}), where $m^2 > 0$. Then $C$ is a bounded
self-adjoint operator on $L^2(M)$. For each $s \in \R$, the
Sobolev space $\Sob_s(M)$ is a real Hilbert space, defined as
completion of $C_c^{\infty}(M)$ in the norm
\bel{SobolevNorm}
    \|f \|^2_s = \< f, C^{-s} f \> .
\ee
The inclusion $\Sob_s \hookrightarrow \Sob_{s+k}$ for $k > 0$ is
Hilbert-Schmidt. Define $\cS := \bigcap_{s <0} \Sob_s(M)$ and
$\cS' := \bigcup_{s>0} \Sob_s(M)$. Then
\[
    \cS
    \ \ \subset \ \
    \Sob_{-1}(M)
    \ \ \subset \ \
    \cS'
\]
form a Gelfand triple, and $\cS$ is a nuclear space.

Recall that $\cS'$ has a natural $\sigma$-algebra of measurable
sets (see for instance \cite{GV,GJ,Simon}). There is a unique
Gaussian probability measure $\mu$ with mean zero and covariance
$C$ defined on the cylinder sets in $\cS'$ (see \cite{GV}).

More generally, one may consider a non-Gaussian, countably-additive measure
$\mu$ on $\cS'$ and the space
\[
    \cE\  := \  L^2(\cS', \mu ).
\]
We are interested in the case that the monomials of the form
$A(\Phi) = \Phi(f_1) \ldots \Phi(f_n)$ for $f_i \in \cS$ are all
elements of $\cE$, and for which their span is dense in $\cE$.
This is of course true if $\mu$ is the Gaussian measure with
covariance $C$.

For an open set $\Om \subset M$, let $\cE_\Om$ denote the closure
in $\cE$ of the set of monomials $A(\Phi) = \prod_i \Phi(f_i)$
where $\supp(f_i) \subset \Om$ for all $i$. Of particular
importance for Euclidean quantum field theory is the positive-time
subspace
\[
    \cE_+ \ :=\  \cE_{\Om_+}\, .
\]

\subsection{The Operator Induced by an Isometry}
\label{sec:operator-induced}

Isometries of the underlying space-time manifold act on a Hilbert space of
classical fields arising in the study of a classical field theory. For $f \in
C^\infty(M)$ and $\psi : M \to M$ an isometry, define
    \[
        f^\psi \ \equiv\  (\psi^{-1})^* f = f \circ \psi^{-1}.
    \]
Since $\det(d\psi) = 1$, the operation $f \to f^\psi$ extends to a
bounded operator on $\Sob_{\pm1}(M)$ or on $L^2(M)$.  A treatment
of isometries for static space-times appears in
\cite{JR:CurvedSp}.

\begin{definition}\label{def:Gamma}
Let $\psi$ be an isometry, and $A(\Phi) = \Phi(f_1) \ldots \Phi(f_n) \in \cE$ a
monomial. Define the induced operator
    \bel{WickOrderedGamma}
        \Gamma(\psi) {A}
        \ \equiv\
        { \Phi({f_1}^\psi) \ldots \Phi({f_n}^\psi) }\,,
    \ee
and extend $\Gamma(\psi)$ by linearity to the domain of
polynomials in the fields, which is dense in $\cE$.
\end{definition}

\section{Osterwalder-Schrader Quantization}
\subsection{Quantization of Vectors (The Hilbert Space $\cH$ of Quantum Theory)}
\label{sec:RP}
In this section we define the quantization map $\cE_+\to\cH$,
where $\cH$ is the Hilbert space of quantum theory. The existence
of the quantization map relies on a condition known as
Osterwalder-Schrader (or reflection) positivity. A  probability
measure $\mu$ on $\cS'$ is said to be \emph{reflection positive}
if
\bel{action}
    \int \overline{ \Gamma(\theta)F}\, F \ d\mu  \geq 0
\ee
for all $F$ in the positive-time subspace $\cE_+\subset \cE$.  Let
$\Th = \Gamma(\theta)$ be the reflection on $\cE$ induced by
$\theta$. Define the sesquilinear form $(A,B)$ on
$\cE_+\times\cE_+$  as $( A, B ) = \< \Th A, B\>_{\cE}$, so
\er{action} states that $(F,F) \geq 0$.

\begin{assumption}[O-S Positivity]
Any measure $d\mu$ that we consider is reflection positive with respect to the
time-reflection $\Theta$.
\end{assumption}

\begin{definition}[OS-Quantization]\label{def:hilbertspace}  Given a
reflection-positive measure $d\mu$, the Hilbert space $\cH$ of quantum theory
is the completion of $\cE_+ /\cN$ with respect to the inner product given by
the sesquilinear form $( A, B )$. Denote the quantization map $\Pi$ for vectors
$\cE_+\to\cH$ by $\Pi(A) = \hat A$,  and write
     \bel{bform}
        \<\hat A, \hat B \>_{\cH}
        = ( A, B )
        = \< \Th A, B\>_{\cE} \quad
        \te{ for } \quad A, B \in \cE_+ \, .
    \ee
\end{definition}

\subsection{Quantization of Operators}\label{sec:general-quant}
The basic quantization theorem gives a sufficient condition to map a (possibly
unbounded) linear operator $T$ on $\cE$ to its quantization, a linear operator
$\hat T$ on $\cH$. Consider a densely-defined operator $T$ on $\cale$, the
unitary time-reflection $\Th$, and the adjoint $T^+ = \Theta T^* \Theta$.  A
preliminary version of the following was also given in \cite{Jaffe:2005a}.

\begin{definition}  [Quantization Condition I]
\label{def:QC-I}
The operator $T$ satisfies QC-I if:
    \begin{itemize}
    \item[i.]  The operator $T$ has a domain $\cD(T)$ dense in $\cE$.
    \item[ii.]  There is a subdomain $\cD_0\subset\cE_+\cap\cD(T)\cap\cD(T^+)$, for which
    $\widehat\cD_0\subset\cH$ is dense.
    \item[iii.]  The transformations $T$ and $T^+$ both map $\cD_0$ into $\cE_+$.
\end{itemize}
\end{definition}

\begin{theorem}[Quantization I] \label{thm:QC-I}
If $T$ satisfies QC-I, then
    \begin{itemize}
    \item[i.] The operators $T\hook\cD_0$ and $T^+\hook\cD_0$ have quantizations
$\hat T$
    and $\widehat{ T^+}$ with domain $\hat\cD_0$.
    \item[ii.] The operators $\hat T^*= \lrp{\hat T\hook{\hat\cD_0} }^*$ and
$\widehat{T^+}$ agree on $\hat\cD_0$.
    \item[iii.] The operator $\hat T\hook\cD_0$ has a
    closure, namely ${\hat T}^{**}$.
    \end{itemize}
\end{theorem}

\begin{proof}
We wish to define the quantization $\hat T$ with the putative domain $\hat
\cD_0$ by
    \bel{OperatorQuantization}
        \hat T\hat A=\widehat{TA}\;.
    \ee
For any vector $A\in\cD_0$ and for any $B\in\lrp{\cD_0\cap\cN}$, it is the case
that $\hat{A}=\widehat{A+B}$.  The transformation $\hat T$ is defined by
\eqref{OperatorQuantization} iff
$\widehat{TA}=\widehat{T(A+B)}=\widehat{TA}+\widehat{TB}$. Hence one needs to
verify that $T:{\cD_0\cap\cN}\to \cN$, which we now do.

The assumption $\cD_0\subset\cD(T^+)$, along with the fact that
$\Theta$ is unitary, ensures that $\Theta\cD_0\subset\cD(T^*)$.  Therefore for
any $F \in \cD_0$,
    \bel{QuantizationAdjoint}
        \< \Th F, T B \>_{\cE}
         = \< T^*\Th F,  B \>_{\cE}
        = \<\Th\lrp{\Th T^* \Th F}, B \>_{\cE}\
        = \<\Th T^+F, B \>_{\cE}\
         = \< \widehat{T^+ F}, \hat B \>_{\cH}\;.
    \ee
In the last step we use the fact assumed in part (iii) of QC-I
that $T^+:\cD_0\to\cE_+$, yielding the inner product of two
vectors in $\cH$. We infer from the Schwarz inequality in $\cH$
that
    \[
        \abs{\< \Th F, T B \>_{\cE}}
        \le \norm{\widehat{T^+ F}}_{\cH}\;\norm{\hat B }_{\cH}
        =0\;.
    \]
As $\< \Th F, T B \>_{\cE}=\<\hat F,\widehat{TB}\>_{\cH}$, this means  that
$\widehat{TB}\perp\hat\cD_0$.  As $\hat\cD_0$ is dense in $\cH$ by QC-I.ii, we
infer $\widehat{TB}=0$. In other words, $T B\in\cN$ as required to define $\hat
T$.

In order show that $\hat\cD_0\subset\cD(\hat T^*)$,  perform a similar
calculation to \eqref{QuantizationAdjoint} with arbitrary $A\in\cD_0$ replacing
$B$, namely
    \bel{Adjoint1}
        \< \hat F, \hat{T}{\hat A} \>_{\cH}
        = \< \Th F, T A \>_{\cE}
        = \<\Th\lrp{\Th T^* \Th F}, A \>_{\cE}\
        = \<\Th{ T^+  F}, A \>_{\cE}\
         = \< \widehat{T^+ F}, \hat A \>_{\cH}\;.
    \ee
The right side is continuous in $\hat A\in\cH$, and therefore  $\hat
F\in\cD(T^*)$.  Furthermore $T^*\hat F=\widehat{T^+F}$. This identity shows
that if $F\in\cN$, then  $\widehat{T^+F}=0$.  Hence $T^+\hook\cD_0$ has a
quantization $\widehat{T^+}$, and we may write \eqref{Adjoint1} as
    \bel{Adjoint2}
        T^*\hat F = \widehat{T^+} \hat F\;,
        \quad\text{for all}\ \
        F\in\cD_0\;.
    \ee
In particular $\hat T^*$ is densely defined so $\hat T$ has a closure.  This
completes the proof.
\end{proof}

\begin{definition}  [Quantization Condition II]
\label{def:QC-II}
The operator $T$ satisfies QC-II if
    \begin{itemize}
        \item[i.]  Both the operator $T$ and its adjoint $T^*$ have dense
domains $\cD(T), \cD(T^*)\subset \cE$.
        \item[ii.] There is a domain $\cD_0\subset\cE_+$ in the common domain of
        $T$, $T^+$, $T^+T$, and $TT^+$.
        \item[iii.] Each operator $T$, $T^+$, $T^+T$, and $TT^+$
        maps $\cD_0$ into $\cE_+$.
    \end{itemize}
\end{definition}

\begin{theorem}[Quantization II] \label{thm:QC-II}
If $T$ satisfies  QC-II, then
    \begin{itemize}
    \item[i.] The operators $T\hook\cD_0$ and $T^+\hook\cD_0$ have quantizations
$\hat T$
    and $\widehat{ T^+}$ with domain $\hat\cD_0$.
    \item[ii.] If $A,B\in\cD_0$, one has
    $\< \hat B, \hat{T}{\hat A} \>_{\cH}=\< \widehat{T^+}\hat B, {\hat A}
\>_{\cH}$.
    \end{itemize}
\end{theorem}

\mbni{Remarks.}
\begin{enumerate}
\item[i.] In Theorem \ref{thm:QC-II} we drop the assumption that the
domain $\hat\cD_0$ is dense, obtaining quantizations $\hat T$ and
$\widehat{T^+}$ whose domains are not necessarily dense. In order
to compensate for this, we assume more properties concerning the
domain  and the range of $T^+$ on $\cE$.

\item[ii.] As $\hat\cD_0$ need not be dense in $\cH$, the adjoint of
$\hat T$ need not be defined.  Nevertheless, one calls the
operator $\hat T$ {\em symmetric} in case one has
    \bel{Symmetric}
         \< \hat B, \hat{T}{\hat A} \>_{\cH}
         = \< \hat{T}\hat B, {\hat A} \>_{\cH}\;,
         \qquad\text{for all}\ \ A,B\in\cD_0\;.
    \ee

\item[iii.] If $\hat S\supset \hat T$ is a densely-defined extension of
$\hat T$, then ${\hat S}^*=\widehat{T^+}$ on the domain
$\hat\cD_0$.
\end{enumerate}

\begin{proof}
We define the quantization $\hat T$ with the putative domain $\hat \cD_0$. As
in the proof of Theorem \ref{thm:QC-I}, this quantization $\hat T$ is
well-defined iff it is the case that $T:{\cD_0\cap\cN}\to \cN$.  For any
$F\in\cD_0\cap\cN$, by definition $\norm{\hat F}_{\cH}=0$.  Also
    \[
        \<TF,TF\>_{\cH}
        = (TF,TF)
        = \lra{\Th T F, TF }_{\cE}
        = \lra{F, T^*\Th TF }_{\cE}\;,
    \]
where one uses the fact that $\cD_0\subset\cD(T^+T)$.  Thus
    \[
        \<TF,TF\>_{\cH}
        = \lra{\Th F, T^+TF }_{\cE}
        = \<F,T^+TF\>_{\cH}\;.
    \]
Here we use the fact that $T^+T$ maps $\cD_0$ to $\cE_+$.  Thus one can use the
Schwarz inequality on $\cH$ to obtain
    \[
        \<TF,TF\>_{\cH}
        \le \norm{\hat{F}}_{\cH} \; \norm{\widehat{T^+TF}}_{\cH}
        =0\;.
    \]
Hence $T:\cD_0\cap\cN\to\cN$, and $T$ has a quantization $\hat T$ with domain
$\hat\cD_0$.

In order verify that $T^+\hook\cD_0$ has a quantization, one needs to show that
$T^+:\cD_0\cap\cN\subset\cN$.  Repeat the argument above with $T^+$ replacing
$T$.  The assumption $TT^+: \cD_0\to \cE_+$ yields for $F\in\cD_0\cap\cN$,
    \[
        \<T^+F,T^+F\>_{\cH}
        = \<T^*\Th F,T^+ F\>_{\cE}
        = \<\Th F, T T^+ F\>_{\cE}
        = \< \hat F, \widehat{TT^+F} \>_{\cH}\;.
    \]
Use the Schwarz inequality in $\cH$ to obtain the desired result that
    \[
        \<T^+F,T^+F\>_{\cH}
        \le \norm{\hat F}_{\cH}\norm{\widehat{TT^+F}}_{\cH}=0\;.
    \]
Hence $T^+$ has a quantization $\widehat{T^+}$ with domain $\hat\cD_0$, and for
$B\in\cD_0$ one has $\widehat{T^+B}=\widehat{T^+}\hat B$. In order to establish
(ii), assume that $A,B\in\cD_0$. Then
    \bea
        \< \hat B, \hat{T}{\hat A} \>_{\cH}
        &=& \< \Th B, T A \>_{\cE}
        = \<\Th\lrp{\Th T^* \Th B}, A \>_{\cE}\
        = \<\Th{ T^+  B}, A \>_{\cE}\nn\\
        &=& \< \widehat{T^+ B}, \hat A \>_{\cH}
        = \<\widehat {T^+} \hat B,\hat A\>_{\cH}\;.
    \eea
\end{proof}
This completes the proof.

\section{Structure and Representation of the Lie Algebra of Killing Fields}

For the remainder of this paper we assume the following, which is
clearly true in the Gaussian case as the Laplacian commutes with
the isometry group $G$. (A further explanation was given in
\cite{JR:CurvedSp}.)

\begin{assumption}
The isometry groups $G$ that we consider leave the measure $d\mu$
invariant, in the sense that $\Ga$, defined above, is a unitary
representation of $G$ on $\cE$.
\end{assumption}

\subsection{The Representation of $\g$ on $\cE$}

\begin{lemma}
Let $G_i$ be an analytic group with Lie algebra $\g_i$ ($i=1,2$),
and let $\la : \g_1 \to \g_2$ be a homomorphism. There cannot
exist more than one analytic homomorphism $\pi : G_1 \to G_2$ for
which $d\pi = \la$. If $G_1$ is simply connected then there is
always one such $\pi$.
\end{lemma}

Let $D = d/dt$ denote the canonical unit vector field on $\R$.
Let $G$ be a real Lie group with algebra $\g$, and let $X \in \g$.
The map $t D \to t X (t \in \R)$ is a homomorphism of $\Lie(\R) \to \g$,
so by the Lemma there is a unique analytic homomorphism $\xi_X : R \to G$
such that $d\xi_X(D) = X$. Conversely, if $\eta$ is an analytic homomorphism
of $\R \to G$, and if we let $X = d\eta(D)$, it is obvious that $\eta = \xi_X$.
Thus $X \mapsto \xi_X$ is a bijection of $\g$ onto the set of analytic
homomorphisms $\R \to G$. The exponential map is defined by $\exp(X) :=
\xi_X(1)$. For complex Lie groups, the same argument applies, replacing
$\R$ with $\C$ throughout.

Since $\g$ is connected, so is $\exp(\g)$. Hence $\exp(\g) \subseteq G^0$,
where $G^0$ denotes the connected component of the identity in $G$.
It need not be the case for a general Lie group that $\exp(\g) = G^0$,
but for a large class of examples (the so-called \emph{exponential groups})
this does hold. For any Lie group, $\exp(\g)$ contains an open neighborhood
of the identity, so the subgroup generated by $\exp(\g)$ always coincides with
$G^0$.

We will apply the above results with $G = \Iso(M)$, the isometry group of $M$,
and $\g =\Lie(G)$ the algebra of global Killing fields. Thus we
have a bijective correspondence between Killing fields and 1-parameter
groups of isometries. This correspondence has a
geometric realization: the 1-parameter group of isometries
\[
    \phi_s = \xi_X(s) = \exp(s X)
\]
corresponding to $X \in \g$ is the flow generated by $X$.

Consider the two different 1-parameter groups of unitary operators:
\begin{enumerate}
\item the unitary group $\phi_s^*$ on $L^2(M)$, and
\item the unitary group $\Ga(\phi_s)$ on $\cE$.
\end{enumerate}
Stone's theorem applies to both of these unitary groups to yield
densely-defined self-adjoint operators on the respective Hilbert spaces.

In the first case, the relevant self-adjoint operator is simply
an extension of $-i X$, viewed as a differential operator
on $C_c^\infty(M)$. This is because for $f \in C_c^\infty(M)$ and $p \in M$,
we have:
\[
    X_p f = (\cL_X f)(p) = \f{d}{ds} f(\phi_s(p)) |_{s=0} .
\]
Thus $-i X$ is a densely-defined symmetric operator on $L^2(M)$,
and Stone's theorem implies that $-i X$ has self-adjoint extensions.

In the second case, the unitary group $\Ga(\phi_s)$ on $\cE$ also
has a self-adjoint generator $\Ga(X)$,  which can be calculated
explicitly. By definition,
\[
    e^{-i s \Ga(X)} \Big[\prod_{i=1}^n \Phi(f_i)\Big]
    =
    \prod_{i=1}^n \Phi(f_i \circ \phi_{-s}) .
\]
Now replace $s \to -s$ and calculate $d/ds |_{s=0}$ applied to both sides of
the last equation to see that
\[
    \Ga(X) \Big[ \prod_{i=1}^n \Phi(f_i) \Big]
    =
    \sum_{j=1}^n \Phi(f_1) \ldots \Phi(-i X f_j) \Phi(f_{j+1}) \ldots
        \Phi(f_n) \, .
\]
One may check that $\Ga$ is a Lie algebra representation of $\g$,
i.e. $\Ga([X,Y]) = [ \Ga(X), \Ga(Y)]$.

\subsection{The Cartan Decomposition of $\g$}

For each $\xi \in \g$, there exists some dense domain
in $\cE$ on which $\Ga(\xi)$ is self-adjoint, as discussed previously.
However, the quantizations $\widehat \Ga(\xi)$ acting on $\cH$ may be
hermitian, anti-hermitian, or neither depending on whether there
holds a relation of the form
\bel{ThetaPM}
    \Ga(\xi) \Theta = \pm  \Theta\Ga(\xi),
\ee
with one of the two possible signs, or whether no such relation
holds.

Even if \er{ThetaPM} holds, to complete the construction of a
unitary representation one must prove that there exists a dense
domain in $\cH$ on which $\widehat \Ga(\xi)$ is self-adjoint or
skew-adjoint. This nontrivial problem will be dealt with in a
later section using Theorems \ref{thm:QC-I} and \ref{thm:QC-II}
and the theory of symmetric local semigroups \cite{KL:81,Fr:80}.
Presently we determine \emph{which} elements within $\g$ satisfy
relations of the form \er{ThetaPM}.

Let $\vth := \th^*$ as an operator on $C^\infty(M)$, and consider
a Killing field $X \in \g$ also as an operator on $C^\infty(M)$.
Define $\cT : \g \to \g$ by
\bel{def-cT}
    \cT(X) := \vth X \vth .
\ee
From \er{def-cT} it is not obvious that the range of $\cT$ is contained in
$\g$. To prove this, we recall some geometric constructions.

Let $M, N$ be
manifolds, let $\psi : M \to N$ be a diffeomorphism, and $X \in \Vect(M)$.
Then
\bel{push-forward}
    \psi^{-1 *} X \psi^* = X(\cdot \circ \psi) \circ \psi^{-1} .
\ee
defines an operator on $C^\infty(N)$. One may check that this
operator is a derivation, thus \er{push-forward} defines a vector
field on $N$. The vector field \er{push-forward} is usually denoted
\[
    \psi_* X = d\psi(X_{\psi^{-1}(p)})
\]
and referred to as the \emph{push-forward} of $X$.

We now wish to show that $\g = \g_+ \oplus \g_-$, where $\g_\pm$
are the $\pm 1$-eigenspaces of $\cT$. This is proven by
introducing an inner product $(X,Y)_\g$ on $\g$ with respect to
which $\cT$ is self-adjoint.

\begin{theorem} \label{main}
Consider $\g$ as a Hilbert space with inner product $(X,Y)_\g$.
The operator $\cT : \g \to \g$ is self-adjoint with $\cT^2 = I$;
hence
    \bel{g-decomp}
        \g = \g_+ \oplus \g_-
    \ee
as an orthogonal direct sum of Hilbert spaces, where $\g_\pm$ are
the $\pm 1$-eigenspaces of $\cT$. Further, $\d_t \in \g_-$ hence
$\dim(\g_-) \geq 1$. Elements of $\g_-$ have hermitian
quantizations, while elements of $\g_+$ have anti-hermitian
quantizations.\footnote{It is not the case that $\g_-$ consists
only of $\d_t$. In particular, $\dim(\g_-) = 2$ for $M = \bH_2$.
It can occur that $\dim \g_+ = 0$.}
\end{theorem}

\begin{proof}
Write \er{def-cT} as
\bel{ct2}
    \cT(X) = \th^{-1 *} X \th^* = \th_* X \, .
\ee
Thus $\cT$ is the operator of push-forward by $\th$. The push-forward
of a Killing field by an isometry is another Killing field, hence
the range of $\cT$ is contained in $\g$. Also, $\cT$ must have a
trivial kernel since $\cT^2 = I$, and this implies that $\cT$ is surjective.
It follows from \er{ct2} that $\cT$ is a Hermitian operator on $\g$.
Hence $\cT$ is diagonalizable and has real eigenvalues which
are square roots of 1. This establishes the decomposition \er{g-decomp}.
That elements of $\g_-$ have hermitian quantizations, while elements of
$\g_+$ have anti-hermitian quantizations follows from Theorem \ref{thm:QC-I}.
\end{proof}

A \emph{Cartan involution} is a Lie algebra homomorphism $\g \to
\g$ which squares to the identity. It follows from \er{def-cT}
that $\cT$ is a Lie algebra homomorphism; thus, Theorem \ref{main}
implies that $\cT$ is a Cartan involution of $\g$. This implies
that the eigenspaces $(\g_+, \g_-)$ form a \emph{Cartan pair},
meaning that
\bel{cartanpair}
    [\g_+,\g_+] \subset \g_+,
    \quad
    [\g_+, \g_-] \subset \g_-,
    \quad \text{and} \quad
    [\g_-, \g_-] \subset \g_+ \, .
\ee
Clearly $\g_+$ is a subalgebra while $\g_-$ is not, and any
subalgebra contained in $\g_-$ is abelian.

\section{Reflection-Invariant and Reflected Isometries}

Let $G = \Iso(M)$ denote the isometry group of $M$, as above. Then
$G$ has a $\Z_2$ subgroup containing $\{ 1, \theta \}$. This
subgroup acts on $G$ by conjugation, which is just the action
$\psi \to \psi^\th := \th \psi \th$. Conjugation is an (inner)
automorphism of the group, so
\[
    (\psi\phi)^\th = \psi^\th \phi^\th,
    \qquad
    (\psi^\th)^{-1} = (\psi^{-1})^\th .
\]

\begin{definition}
We say that $\psi \in G$ is {\bf reflection-invariant} if
    \[
    \psi^\th = \psi,
    \]
and that $\psi$ is {\bf reflected} if
    \[
    \psi^\th = \psi^{-1}.
    \]
Let $G_{RI}$ denote the subgroup of $G$ consisting
of reflection-invariant elements, and let $G_R$ denote the subset
of reflected elements.
\end{definition}

Note that $G_{RI}$ is the stabilizer of the $\Z_2$ action, hence a
subgroup. An alternate proof of this proceeds using $G_{RI} =
\exp(\g_+)$. Although $G_R$ is closed under the taking of inverses
and does contain the identity, the product of two reflected
isometries is no longer reflected unless they commute. Generally,
the product of an element of $G_R$ with an element of $G_{RI}$ is
neither an element of $G_R$ nor of $G_{RI}$. The only isometry
that is both reflection-invariant and reflected is $\th$ itself.
Thus we have:
\[
    G_R \cap G_{RI}
    \ = \
    \{ 1, \th \}
    \ \subset\
    G_R \cup G_{RI}
    \ \subsetneq\
     G .
\]

\begin{theorem} \label{thm:generation}
Let $G^0$ denote the connected component of the identity in $G$.
Then $G^0$ is generated by $G_R \cup G_{RI}$. (This is a form of
the Cartan decomposition for $G$.)
\end{theorem}

\begin{proof}
Since $\g = \g_+ \oplus \g_-$ as a direct sum of vector spaces
(though not of Lie algebras), we have
\[
    G^0 = \lrabig{ \!\exp(\g) }
    =
    \lrabig{ \!\exp(\g_+) \cup \exp(\g_-) } .
\]
Choose bases
$
    \{ \xi_{\pm, i} \}_{i = 1, \ldots, n_\pm}
    \text{ for }
    \g_\pm
$
respectively. Then we have:
\[
    G^0 = \lrabig{
    \{ \exp(s \xi_{+, i}) : 1 \leq i \leq n_+,\, s \in \R \}
    \cup
    \{ \exp(s \xi_{-, j}) : 1 \leq j \leq n_-,\, s \in \R \}
    } .
\]
Furthermore, $\exp(s \xi_{-, i})$ is reflected, while $\exp(s
\xi_{+, i})$ is reflection-invariant, completing the proof.
\end{proof}

\begin{corollary}
The Lie algebra of the subgroup $G_{RI}$ is $\g_+$.
\end{corollary}

To summarize, the isometry group of a static space-time can always
be generated by a collection of $n$ ($=\dim \g$) one-parameter
subgroups, each of which consists either of reflected isometries,
or reflection-invariant isometries.


\section{Construction of Unitary Representations}


\subsection{Self-adjointness of Semigroups}

In this section, we recall several known results on
self-adjointness of semigroups. Roughly speaking, these results
imply that if a one-parameter family $S_\al$ of unbounded
symmetric operators satisfies a semigroup condition of the form
$S_\al S_\beta = S_{\al + \beta}$, then under suitable conditions
one may conclude essential self-adjointness.

A theorem of this type appeared in a 1970 paper of Nussbaum
\cite{Nussbaum}, who assumed that the semigroup operators have a
common dense domain. The result was rediscovered independently by
Fr\"ohlich, who applied it to quantum field theory in several
important papers \cite{Fr:76,DF:77}. For our intended application
to quantum field theory, it turns out to be very convenient to
drop the assumption that $\exists \, a$ such that the $S_\al$ all
have a common dense domain for $|\al| < a$, in favor of the weaker
assumption that $\bigcup_{\al > 0} D(S_\al)$ is dense.

A generalization of Nussbaum's theorem which allows the domains of
the semigroup operators to vary with the parameter, and which only
requires the \emph{union} of the domains to be dense, was later
formulated and two independent proofs were given: one by
Fr\"ohlich \cite{Fr:80}, and another by Klein and Landau
\cite{KL:81}. The latter also used this theorem in their
construction of representations of the Euclidean group and the
corresponding analytic continuation to the Lorentz group
\cite{KL:87}.

In order to keep the present article self-contained, we first
define symmetric local semigroups and then recall the refined
self-adjointness theorem of Fr\"ohlich, and Klein and Landau.

\begin{definition} \label{def:SymmLocalSemi}
Let $\cH$ be a Hilbert space, let $T > 0$ and for each $\al \in
[0,T]$, let $S_\al$ be a symmetric linear operator on the domain
$\cD_\al \subset \cH$, such that:
\begin{enumerate}
\item[{\rm (i)}]  $\cD_\al \supset \cD_\beta$ if $\al \leq \beta$ and
$
    \cD := \bigcup_{0 < \al \leq T} \cD_\al
    \ \te{ is dense in } \ \cH ,$
\item[{\rm (ii)}] $\al \to S_\al$ is weakly continuous,
\item[{\rm (iii)}] $S_0 = I$, $S_\beta(\cD_\al) \subset \cD_{\al-\beta}$ for $0
\leq \beta \leq \al \leq T$, and
\item[{\rm (iv)}] $S_\al S_\beta = S_{\al+\beta}$ on $\cD_{\al+\beta}$ for
$\al,\beta,\al+\beta \in [0,T]$.
\end{enumerate}
In this situation, we say that $(S_\al, \cD_\al, T)$ is a
\emph{symmetric local semigroup}.
\end{definition}

It is important that $\cD_\al$ is \emph{not} required to be dense
in $\cH$ for each $\al$; the only density requirement is (i).

\def\FootNoteText{The authors of \cite{Fr:80,KL:81} also showed that
\[
    \Dhat := \bigcup_{0 < \al \leq S}
    \Big[ \bigcup_{0 < \beta < \al} S_\beta(\cD_\al) \Big],
    \quad \te{ where } \quad 0 < S \leq T ,
\]
is a \emph{core} for $A$, i.e. $(A, \Dhat)$ is essentially self-adjoint.}

\begin{theorem}[\cite{KL:81,Fr:80}] \label{thm:SLS}
For each symmetric local semigroup $(S_\al, \cD_\al, T)$, there
exists a unique self-adjoint operator $A$ such that\footnote{\FootNoteText}
\[
    \cD_\al \subset D(e^{-\al A}) \ \text{ and } \
    S_\al = e^{-\al A}|_{\cD_\al} \ \text{ for all }
    \al \in [0,T] .
\]
Also, $A \geq -c$ if and only if $\norm{S_\al f} \leq e^{c \al} \norm{f}$
for all $f \in \cD_\al$ and $0 < \al < T$.
\end{theorem}

\subsection{Reflection-Invariant Isometries} \label{sec:RI=>PTI}

\begin{lemma} \label{lemma:RI=>PTI}
Let $\psi$ be a reflection-invariant isometry and assume $\exists
\, p \in \Om_+$ such that $\psi(p) \in \Om_+$. Then $\psi$
preserves the positive-time subspace, i.e. $\psi(\Om_+) \subseteq
\Om_+$.
\end{lemma}

\begin{proof}
We first prove that $\psi(\Sigma) \subseteq \Sigma$. Suppose not;
then $\exists \, p \in \Sigma$ with $\psi(p) \not\in \Sigma$.
Assume $\psi(p) \in \Om_+$ (without loss of generality: we could
repeat the same argument with $\psi(p) \in\Om_-$). Then $\Om_+$
contains $(\theta \psi \theta)(p) = \theta\psi(p) \in \Om_-$, a
contradiction since $\Om_- \cap \Om_+ = \emptyset$. We used the
fact that $\theta|_\Sigma = \text{id}$ so $\th(p) = p$. Hence
$\psi$ restricts to an isometry of $\Sigma$. It follows that the
restriction of $\psi$ to $M' = M \setminus \Sigma$ is also an
isometry. However, $M' = \Om_- \sqcup \Om_+$, where $\sqcup$
denotes the disjoint union. Therefore $\psi(\Om_+)$ is wholly
contained in either $\Om_+$ or $\Om_-$, since $\psi$ is a
homeomorphism and so $\psi(\Om_+)$ is connected. The possibility
that $\psi(\Om_+) \subseteq \Om_-$ is ruled out by our assumption,
so $\psi(\Om_+) \subseteq \Om_+$.
\end{proof}

Lemma \ref{lemma:RI=>PTI} has the immediate consequence that if
$\xi \in \g_+$ then the one-parameter group associated to $\xi$ is
positive-time-invariant. This result plays a key role in the proof
of Theorem \ref{thm:Generators}.

\subsection{Construction of Unitary Representations}

The rest of this section is devoted to proving that the theory of
symmetric local semigroups can be applied to the quantized
operators on $\cH$ corresponding to each of a set of 1-parameter
subgroups of $G = \Iso(M)$. The proof relies upon Lemma
\ref{lemma:RI=>PTI}, and Theorems \ref{thm:QC-I}, \ref{thm:QC-II}
and \ref{thm:SLS}.

\begin{theorem} \label{thm:Generators}
Let $(M, g_{ab})$ be a quantizable static space-time. Let $\xi$ be a Killing
field which lies in $\g_+$ or $\g_-$, with associated one-parameter group of
isometries $\{ \phi_\al \}_{\al \in \R}$. Then there exists a densely-defined
self-adjoint operator $A_\xi$ on $\cH$ such that
\[
    \widehat{\Ga}(\phi_\al) =
    \begin{cases}
    e^{-\al A_\xi}, & \te{ if } \ \xi \in \g_- \\
    e^{i \al A_\xi} & \te{ if } \ \xi \in \g_+ .
    \end{cases}
\]
\end{theorem}

\begin{proof}
First suppose that $\xi \in \g_-$, which implies that the
isometries $\phi_\al$ are reflected, and so $\Ga(\phi_\al)^+ =
\Ga(\phi_\al)$. Define
\[
    \Om_{\xi,\al} := \phi_\al^{-1}(\Om_+) .
\]
For all $\al$ in some neighborhood of zero, $\Om_{\xi,\al}$ is a
nonempty open subset of $\Om_+$, and moreover, as $\al \to 0^+$,
$\Om_{\xi,\al}$ increases to fill $\Om_+$ with $\Om_{\xi,0} =
\Om_+$. These statements follow immediately from the fact that,
for each $p \in \Om_+$, $\phi_\al(p)$ is continuous with respect
to $\al$, and $\phi_0$ is the identity map.

Since $\phi_\al(\Om_{\xi,\al}) \subseteq \Om_+$, we infer that
$\Ga(\phi_\al) \cE_{\Om_{\xi,\al}} \subseteq \cE_+$. By Theorem
\ref{thm:QC-II}, $\Ga(\phi_\al)$ has a quantization which is a
symmetric operator on the domain
\[
    \cD_{\xi,\al} := \Pi(\cE_{\Om_{\xi,\al}}).
\]
Note that $\cD_{\xi,\al}$ is not necessarily dense in $\cH$.
\footnote{Density of $\cD_{\xi,\al}$ would be implied by a
Reeh-Schlieder theorem, which we do not prove except in the free
case. Theorem \ref{thm:SLS} removes the need for a Reeh-Schlieder
theorem in this argument.} We now show that Theorem \ref{thm:SLS}
can be applied.

Fix some positive constant $a$ with $\Om_{\xi,a}$ nonempty.
Note that
\[
    \bigcup_{0 < \al \leq a} \Om_{\xi,\al}
    =
    \Om_+
    \quad \Rightarrow \quad
    \bigcup_{0 < \al \leq a} \cE_{\Om_{\xi,\al}}
    =
    \cE_+.
\]
It follows that
    \[
    \cD_\xi := \bigcup_{0 < \al \leq a} \cD_{\xi,\al}
    \]
is dense in $\cH$. This establishes condition (i) of Definition
\ref{def:SymmLocalSemi}, and the other conditions are routine
verifications. Theorem \ref{thm:SLS} implies existence of a
densely-defined self-adjoint operator $A_\xi$ on $\cH$, such that
\[
    \widehat{\Ga}(\phi_\al) = \exp(-\al A_\xi)
    \ \te{ for all } \
    \al \in [0,a] \, .
\]
This proves the theorem in case $\xi \in \g_-$.

Now suppose that $\xi \in \g_+$, implying that the isometries
$\phi_\al$ are reflection-invariant, and
\[
    \Ga(\phi_\al)^+ = \Ga(\phi_\al)^{-1} = \Ga(\phi_{-\al})
    \ \te{ on } \
    \cE.
\]
Lemma \ref{lemma:RI=>PTI} implies that $\Ga(\phi_\al)\cE_+
\subseteq \cE_+$. By Theorem \ref{thm:QC-I}, $\Ga(\phi_\al)$ has a
quantization $\widehat{\Ga}(\phi_\al)$ which is defined and
satisfies
\[
    \widehat{\Ga}(\phi_\al)^* = \widehat{\Ga}(\phi_\al)^{-1}
\]
on the domain $\Pi(\cE_+)$, which is dense in $\cH$ by definition.
In this case we do not need Theorem \ref{thm:SLS}; for each $\al$,
$\widehat{\Ga}(\phi_\al)$ extends by continuity to a one-parameter
unitary group defined on all of $\cH$ (not only for a dense
subspace). By Stone's theorem,
\[
    \widehat{\Ga}(\phi_\al) = \exp(i \al A_\xi)
\]
for $A_\xi$ self-adjoint and for all $\al \in \R$.
The proof is complete.
\end{proof}


\section{Analytic Continuation}

Each Riemannian static space-time $(M,g_{ab})$ has a Lorentzian
continuation $\Lor{M}$, which we construct as follows. In adapted
coordinates, the metric $g_{ab}$ on $M$ takes the form
\bel{smetric}
    ds^2 = \cF(x) dt^2 + \cG_{\mu\nu}(x) dx^\mu dx^\nu .
\ee
The analytic continuation $t \to -i t$ of \er{smetric} is standard
and gives a metric of Lorentz signature, $\Lor{ds^2} = -\cF\, dt^2
+ \cG \, dx^2$, by which we define the Lorentzian space-time
$\Lor{M}$. Einstein's equation $\mathrm{Ric}_g = k \, g$ is
preserved by the analytic continuation, but we do not use this
fact anywhere in the present paper.

Let $\{ \xi_i^{(\pm)} : 1 \leq i \leq n_\pm \}$ be bases of
$\g_\pm$, respectively. Let $A_i^{(\pm)} = A_{\xi_i^{(\pm)}}$ be
the densely-defined self-adjoint operators on $\cH$, constructed
by Theorem \ref{thm:Generators}. Let
\bel{UnitaryGroups}
    U_i^{(\pm)}(\al) = \exp(i \al A_i^{(\pm)}) \, , \
    \te{ for } \
    1 \leq i \leq n_\pm
\ee
be the associated one-parameter unitary groups on $\cH$.

We claim that the group generated by the $n = n_+ + n_-$
one-parameter unitary groups \er{UnitaryGroups} is isomorphic to
the identity component of
    \[
    \Lor{G} := \Iso(\Lor{M}),
    \]
the group of Lorentzian isometries. Since locally, the group
structure is determined by its Lie algebra, it suffices to check
that the generators satisfy the defining relations of $\Lor{\g} :=
\Lie(\Lor{G})$.

Since quantization of operators preserves multiplication, we have
\bel{pres-mult}
    X,Y,Z \in \g, \ [X,Y]=Z \quad \Rightarrow \quad
    [\widehat \Ga(X), \widehat \Ga(Y)] = \widehat \Ga(Z) .
\ee
In what follows, we will use the notation $\widehat{\g}_\pm$ for
$\{ \widehat\Ga(X) : X \in \g_\pm \}$.

Quantization converts the elements of $\g_-$ from skew operators
into Hermitian operators; i.e. elements of $\widehat{\g}_-$ are
Hermitian on $\cH$ and hence, elements of $i \, \widehat{\g}_-$
are skew-symmetric on $\cH$. Thus $\widehat{\g}_+ \oplus i \,
\widehat{\g}_-$ is a Lie algebra represented by skew-symmetric
operators on $\cH$.

\begin{theorem} \label{IsomorphismTheorem}
We have an isomorphism of Lie algebras:
    \bel{Isomorphism}
        \Lor{\g}
        \ \cong\
        \widehat{\g}_+ \oplus i \, \widehat{\g}_- \, .
    \ee
\end{theorem}

\begin{proof}
Let $M_\C$ be the manifold obtained by allowing the $t$ coordinate
to take values in $\C$. Define $\psi : M_\C \to M_\C$ by $t \mapsto -i t$.
Then $\Lor{\g}$ is generated by
\[
    \{ \xi_i^{(+)} \}_{1 \leq i \leq n_+}
    \cup \{ \eta_j \}_{1 \leq j \leq n_- },
    \quad
    \te{ where } \quad
    \eta_j := i \psi^*\big(\xi_j^{(-)}\big) .
\]

It is possible to define a set of real structure constants
$f_{ijk}$ such that
\bel{StructConst}
    [\xi_i^{(-)}, \xi_j^{(-)}] = \sum_{k=1}^{n_+} f_{ijk} \xi_k^{(+)} \, .
\ee
Applying $\psi^*$ to both sides of \er{StructConst}, the
commutation relations of $\Lor{\g}$ are seen to be
\bel{g-lor-relns}
    [\eta_i, \eta_j] = - f_{ijk} \xi_k^{(+)},
\ee
together with the same relations for $\g_+$ as before. Now
\er{pres-mult} implies that \er{g-lor-relns} are the precisely the
commutation relations of $\widehat{\g}_+ \oplus i \,
\widehat{\g}_-$, completing the proof of \er{Isomorphism}.
\end{proof}

\begin{corollary} \label{Cor:MainQuantization}
Let $(M, g_{ab})$ be a quantizable static space-time. The unitary
groups \er{UnitaryGroups} determine a unitary representation of
$\Lor{G^0}$ on $\cH$.
\end{corollary}

\subsection{Conclusions}

We have obtained the following conclusions. There is a unitary
representation of the group $\Lor{G^0}$ on the physical Hilbert
space $\cH$ of quantum field theory on the static space-time $M$.
This representation maps the time-translation subgroup into the
unitary group $\exp(it H)$, where the energy $H \geq 0$ is a
positive, densely-defined self-adjoint operator corresponding to
the Hamiltonian of the theory. The Hilbert space $\cH$ contains a
ground state $\Psi_0 = \hat 1$ which is such that $H\Psi_0 = 0$
and $\Psi_0$ is invariant under the action of all spacetime
symmetries. We obtain these results via analytic continuation from
the Euclidean path integral, under mild assumptions on the measure
which should include all physically interesting examples. This is
done without introducing the field operators; nonetheless,
Theorems \ref{thm:QC-I} and \ref{thm:QC-II} do suffice to
construct field operators. In the special case $M = \R^d$ with $G
= \SO(4)$, we obtain a unitary representation of the proper
orthochronous Lorentz group, $\Lor{G^0} = L_+^\uparrow =
\SO^0(3,1)$.


\section{Hyperbolic Space and Anti-de Sitter Space}

Consider Euclidean quantum field theory on $M = \bH^d$. The metric
is
\[
    ds^2 = r^{-2} \sum_{i=1}^{d} dx_i^2,
\]
where we define $r = x^{d}$ for convenience. The Laplacian is
\bel{eq:hyp-lapl}
    \De_{\bH^d} = (2-d) r \f{\d}{\d r}
    +
    r^2 \De_{\R^d} \, .
\ee The $d-1$ coordinate vector fields $\{ \d / \d x^i : i \ne d \}$ are all
static Killing fields, and any one of the coordinates $x^i \ (i \ne d)$ is a
satisfactory representation of time in this space-time. It is convenient to
define $t = x^1$ as before, and to identify $t$ with time.

The time-zero slice is $M_0 = \bH^{d-1}$. From
\[
    \bH^d = \{ v \in \R^{d,1} \mid \<v,v\> = -1,\, v_0 > 0\}
\]
it follows that $\Isom(\bH^d) = O^+(d,1)$ and the
orientation-preserving isometry group is $SO^+(d,1)$.

For constant curvature spaces, one may solve Killing's equation $\cL_K g = 0$
explicitly. Let us illustrate the solutions and their quantizations for $d=2$.
The three Killing fields
\bel{H2Fields}
    \xi = \d_t, \quad
    \eta = t \d_t + r \d_r, \quad
    \zeta = (t^2 - r^2) \d_t + 2tr \, \d_r
\ee
are a convenient basis for $\g$. Any $d$-dimensional
manifold satisfies $\dim \g \leq d(d+1)/2$, manifolds
saturating the bound are said to be \emph{maximally symmetric},
and $\bH^d$ is maximally symmetric.

Now, $\d_t f(-t) = -f'(-t)$ so $\d_t\Theta = -\Theta \d_t$, hence
$\d_t \in \g_-$. Similar calculations show $[\Theta, \eta] = 0$
and $\Theta \zeta = - \zeta \Theta$. Thus $\eta$ spans $\g_+$,
while $\d_t, \zeta$ span $\g_-$. The commutation
relations\footnote{Note that quite generally $[\g_-, \g_-] \subseteq \g_+$ so
it's automatic that $[\zeta, \d_t]$ is proportional to $\eta$.} for
$\g$ are:
\[
    [\eta, \zeta] = \zeta, \quad
    [\eta, \d_t] = - \d_t, \quad
    [\zeta, \d_t] = -2 \eta .
\]
These calculations verify that $(\g_+, \g_-)$ forms a Cartan pair,
as defined in \er{cartanpair}.

\begin{figure}
    \includegraphics{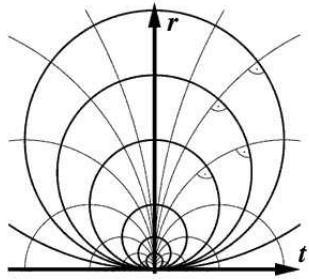}
    \centering
    \caption{\label{fig:hyper}Flow lines of the Killing field
    $\zeta = (t^2 - r^2) \d_t + 2tr \, \d_r$ on $\bH^d$.}
\end{figure}

The flows associated to \er{H2Fields} are easily visualized: $\xi$
is a right-translation, and $\eta$ flow-lines are radially outward
from the Euclidean origin. The flows of $\zeta$ are Euclidean
circles, indicated by the darker lines in Figure \ref{fig:hyper}.
Hence the flows of $\eta$ are defined on all of $\cE_+$, while the
flows of $\zeta$ are analogous to space-time rotations in $\R^2$,
and hence, must be defined on a wedge of the form
    \[
        W_\al = \{ (t,r) \ :\ t,r > 0,\ \tan^{-1}(r/t) < \al\}.
    \]
The simple geometric idea of Section \ref{sec:RI=>PTI} is nicely confirmed
in this case: the flows of $\eta$ (the generator of $\g_+$)
preserve the $t=0$ plane, and are separately isometries of $\Om_+$ and
$\Om_-$.

Corollary \ref{Cor:MainQuantization} implies that the procedure
outlined above defines a unitary representation of the identity
component of $\Iso(AdS_2)$ on the physical Hilbert space $\cH$ for
quantum field theory on this background, including theories with
interactions that preserve the symmetry. Since $\Iso(AdS_{d+1}) =
\SO(d,2)$, we have a unitary representation of $\SO^0(1,2)$. The
latter is a noncompact, semisimple real Lie group, and thus it has
no finite-dimensional unitary representations, but a host of
interesting infinite-dimensional ones.

\appendix

\section{Euclidean Reeh-Schlieder Theorem}

We prove the Euclidean Reeh-Schlieder property for free
theories on curved backgrounds. It is reasonable to expect
this property to extend to \emph{interacting} theories on curved
backgrounds, but it would have to be established for
each such model since it depends explicitly on the two-point
function.

The Reeh-Schlieder theorem guarantees the existence of a dense quantization
domain based on any open subset of $\Om_+$. For this reason, one could use the
Reeh-Schlieder (RS) theorem with Nussbaum's theorem \cite{Nussbaum} to
construct a second proof of Theorem \ref{thm:Generators} under the additional
assumption that $M$ is real-analytic.

Fortunately, our proof of Theorem \ref{thm:Generators}
is completely independent of the Reeh-Schlieder property.
This has two advantages: we do not have to assume $M$ is a
real-analytic manifold and, more importantly, our proof
of Theorem \ref{thm:Generators} generalizes immediately
and transparently to interacting theories as long as the
Hilbert space $\cH$ is not modified by the interaction.

We state and prove this using the one-particle space; however,
the result clearly extends to the quantum-field Hilbert space.

\begin{theorem} \label{thm:ReehSchlieder}
Let $M$ be a quantizable static space-time endowed with a real-analytic
structure, and assume that $g_{ab}$ is real-analytic. Let $\cO \subset \Om_+$
and $\cD = C^\infty(\cO) \subset L^2(\Om_+)$. Then $\widehat{\cD}^\perp = \{ 0
\}$.
\end{theorem}

\begin{proof}
Let $f \in L^2(\Om_+)$ with $\hat f \perp \cD$.
For $x \in \Om_+$, define
\[
    \eta(x) := \< \hat f, \hat \delta_x\>_{\cH}
        =
    \< \Th f, C \de_x \>_{L^2} .
\]
Real-analyticity of $\eta(x)$ follows from the real-analyticity
of (the integral kernel of) $C$, which in turn follows from
the elliptic regularity theorem in the real-analytic category
(see for instance  \cite[Sec.~II.1.3]{Bers}).
Now by assumption, for any $g \in C_c^\infty(\cO)$, we have
\[
    0 = \< \hat g, \hat f\>_{\cH} = \< \Th f, Cg \>_{L^2(M)} .
\]
Let $g \to \de_x$ for $x \in \cO$. Then
$
    0 = \< \Th f, C \de_x \>_{L^2} \equiv \eta(x) .
$
Since $\eta|_{\cO} = 0$, by real-analyticity we infer the
vanishing of $\eta$ on $\Om_+$, completing the proof.
\end{proof}

\subsection*{Acknowledgements}
We are grateful to Hanno Gottschalk and Alexander Strohmaier for
helpful discussions, and G.R. is grateful to the Universit\"at
Bonn for their hospitality during February 2007.

\bibliographystyle{plain}

\begin{thebibliography}{10}

\bibitem{Bers}
Lipman Bers, Fritz John, and Martin Schechter.
\newblock {\em Partial differential equations}.
\newblock American Mathematical Society, Providence, R.I., 1979.
\newblock Lectures in Applied Mathematics 3.

\bibitem{BirrellDavies}
N.~D. Birrell and P.~C.~W. Davies.
\newblock {\em Quantum fields in curved space}, volume~7 of {\em Cambridge
  Monographs on Mathematical Physics}.
\newblock Cambridge University Press, Cambridge, 1982.

\bibitem{DF:77}
W.~Driessler and J.~Fr{\"o}hlich.
\newblock The reconstruction of local observable algebras from the {E}uclidean
  {G}reen's functions of relativistic quantum field theory.
\newblock {\em Ann. Inst. H. Poincar\'e Sect. A (N.S.)}, 27(3):221--236, 1977.

\bibitem{Fr:80}
J.~Fr{\"o}hlich.
\newblock Unbounded, symmetric semigroups on a separable {H}ilbert space are
  essentially selfadjoint.
\newblock {\em Adv. in Appl. Math.}, 1(3):237--256, 1980.

\bibitem{Fr:76}
J{\"u}rg Fr{\"o}hlich.
\newblock The pure phases, the irreducible quantum fields, and dynamical
  symmetry breaking in {S}ymanzik-{N}elson positive quantum field theories.
\newblock {\em Ann. Physics}, 97(1):1--54, 1976.

\bibitem{Fulling}
Stephen~A. Fulling.
\newblock {\em Aspects of quantum field theory in curved space-time}, volume~17
  of {\em London Mathematical Society Student Texts}.
\newblock Cambridge University Press, Cambridge, 1989.

\bibitem{GV}
I.~M. Gel{\cprime}fand and N.~Ya. Vilenkin.
\newblock {\em Generalized functions. {V}ol. 4}.
\newblock Academic Press [Harcourt Brace Jovanovich Publishers], New York, 1964
  [1977].
\newblock Applications of harmonic analysis, Translated from the Russian by
  Amiel Feinstein.

\bibitem{GJ}
James Glimm and Arthur Jaffe.
\newblock {\em Quantum physics}.
\newblock Springer-Verlag, New York, second edition, 1987.
\newblock A functional integral point of view.

\bibitem{Jaffe:2000a}
Arthur Jaffe.
\newblock Constructive quantum field theory.
\newblock In {\em Mathematical physics 2000}, pages 111--127. Imp. Coll. Press,
  London, 2000.

\bibitem{Jaffe:2005a}
Arthur Jaffe.
\newblock {\em Introduction to Quantum Field Theory}.
\newblock 2005.
\newblock Lecture notes from Harvard Physics 289r, available in part online
  at\\ {\tt\small http://www.arthurjaffe.com/Assets/pdf/IntroQFT.pdf}.

\bibitem{JR:CurvedSp}
Arthur Jaffe and Gordon Ritter.
\newblock Quantum field theory on curved backgrounds. i. the euclidean
  functional integral.
\newblock {\em Comm. Math. Phys.}, 270(2):545--572, 2007.

\bibitem{KL:81}
Abel Klein and Lawrence~J. Landau.
\newblock Construction of a unique selfadjoint generator for a symmetric local
  semigroup.
\newblock {\em J. Funct. Anal.}, 44(2):121--137, 1981.

\bibitem{KL:87}
Abel Klein and Lawrence~J. Landau.
\newblock From the {E}uclidean group to the {P}oincar\'e group via
  {O}sterwalder-{S}chrader positivity.
\newblock {\em Comm. Math. Phys.}, 87(4):469--484, 1983.

\bibitem{Nussbaum}
A.~E. Nussbaum.
\newblock Spectral representation of certain one-parametric families of
  symmetric operators in {H}ilbert space.
\newblock {\em Trans. Amer. Math. Soc.}, 152:419--429, 1970.

\bibitem{OS1}
Konrad Osterwalder and Robert Schrader.
\newblock Axioms for {E}uclidean {G}reen's functions.
\newblock {\em Comm. Math. Phys.}, 31:83--112, 1973.

\bibitem{OS2}
Konrad Osterwalder and Robert Schrader.
\newblock Axioms for {E}uclidean {G}reen's functions. {II}.
\newblock {\em Comm. Math. Phys.}, 42:281--305, 1975.
\newblock With an appendix by Stephen Summers.

\bibitem{Simon}
Barry Simon.
\newblock {\em The {$P(\phi )\sb{2}$} {E}uclidean (quantum) field theory}.
\newblock Princeton University Press, Princeton, N.J., 1974.
\newblock Princeton Series in Physics.

\bibitem{WaldSurvey}
Robert~M. Wald.
\newblock Quantum field theory in curved space-time.
\newblock In {\em Gravitation et quantifications (Les Houches, 1992)}, pages
  63--167. North-Holland, Amsterdam, 1995.

\end{thebibliography}
\def\cprime{$'$}

\end{document}